\documentclass[twocolumn,showpacs,preprintnumbers,amsmath,amssymb]{revtex4-1}
\usepackage{graphicx}
\usepackage{color}

\begin{document}
\title{Coupling Tension and Shear for Highly Sensitive Graphene-Based Strain Sensors}
\author{Zenan Qi$^{1}$,Jian Zhang$^{2}$,Guiping Zhang$^{3}$ and Harold S. Park$^{1}$}
\address{$^1$Department of Mechanical Engineering, Boston University,
  Boston, MA 02215}
\address{$^2$ Microsoft Corporation 15700 NE 39th St Redmond, WA 98052}
\address{$^3$Department of Physics, Renmin University of China,
Beijing 100872, China}
\date{\today}

\begin{abstract}

We report, based on its variation in electronic transport to coupled tension and shear deformation, a highly sensitive graphene-based strain sensor consisting of an armchair graphene nanoribbon (AGNR) between metallic contacts.  As the nominal strain at any direction increases from 2.5 to 10\%, the conductance decreases, particularly when the system changes from the electrically neutral region. At finite bias voltage, both the raw conductance and the relative proportion of the conductance depends smoothly on the gate voltage with negligible fluctuations, which is in contrast to that of pristine graphene.  Specifically, when the nominal strain is 10\% and the angle varies from $0^{\circ}$ to $90^{\circ}$, the relative proportion of the conductance changes from 60 to $\sim$90\%.
\end{abstract}
\maketitle

Graphene has been proposed for many applications due to its unique physical properties \cite{graphene1,graphene2,graphene3,gas-detector}, in which the electronic transport through graphene nanoribbon would be affected by line defect \cite{R1}. Of specific interest to the present work, it has been proposed as a strain sensor due to change in the conductivity of graphene-based materials \cite{strain-e1,strain-e2,strain-e3,strain-e4,strain-e5,strain-simulation1,strain-simulation2}. The field of graphene-based strain sensing has rapidly developed since the experimental observation of the increase in resistance of CVD graphene samples when strain is applied in the direction of the electrical current \cite{strain-e1}. In order to accurately determine the direction and magnitude of strain, a triaxial graphene-based strain sensor composite was proposed; it was found that the resistance of graphene may be enhanced or reduced by the strain in certain directions \cite{strain-e2}.  The sensitivity of graphene to strain originates from the deformation of carbon-carbon bonds, which alter the hopping integrals, and thus the electronic transport in graphene.

Though the effect of strain on the band structure of graphene and narrow graphene nanoribbons (GNRs) has been widely discussed \cite{strain-TB, strain-band structure1, strain-band structure2, strain-band structure3,R2}, strain sensing based on electronic transport through graphene and GNRs has recently become of wide interest \cite{strain-transport1,strain-transport2,strain-transport3,strain-transport4,strain-transport-GNR1,strain-transport-GNR2,strain-transport5}. In the present work, motivated by recent experimental findings of the potential benefits of coupled tension/shear deformation to simulate the strain generated due to the movement of human fingers \cite{strain-e2}, we theoretically study coupled tension and shear deformation on the transport through armchair graphene nanoribbons between metallic contacts.

\begin{figure}
  \centering
  \includegraphics[scale=0.40]{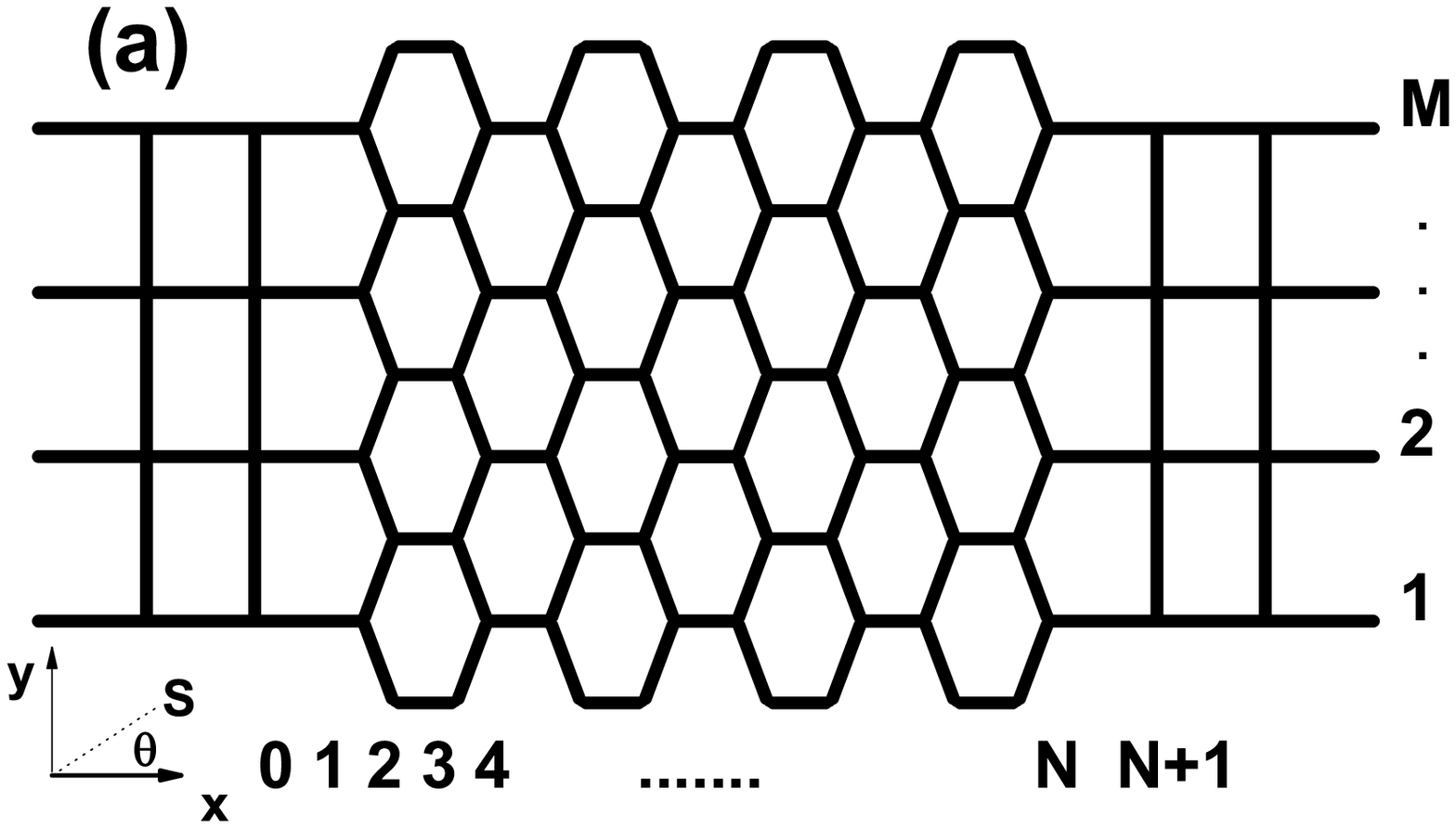}
  \includegraphics[scale=0.3]{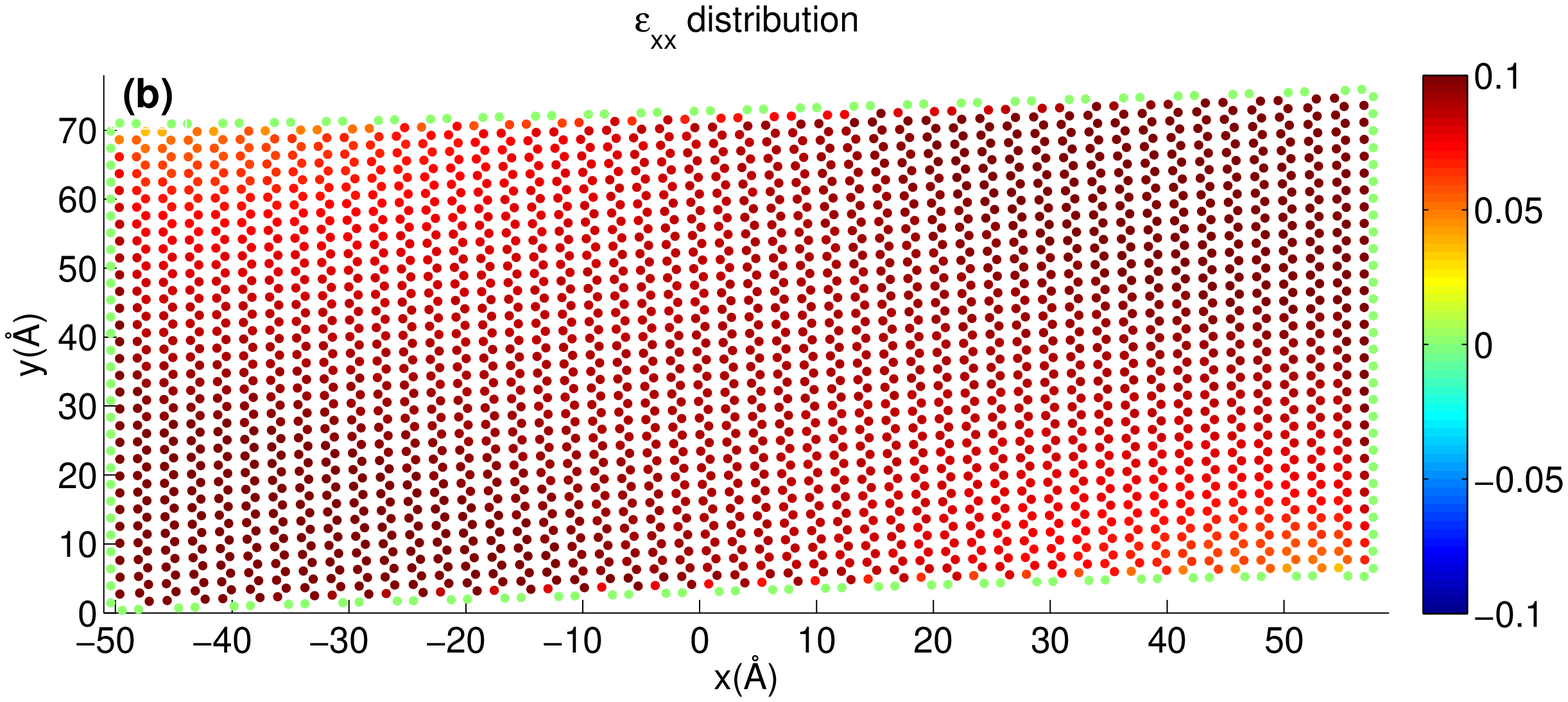}
  \includegraphics[scale=0.3]{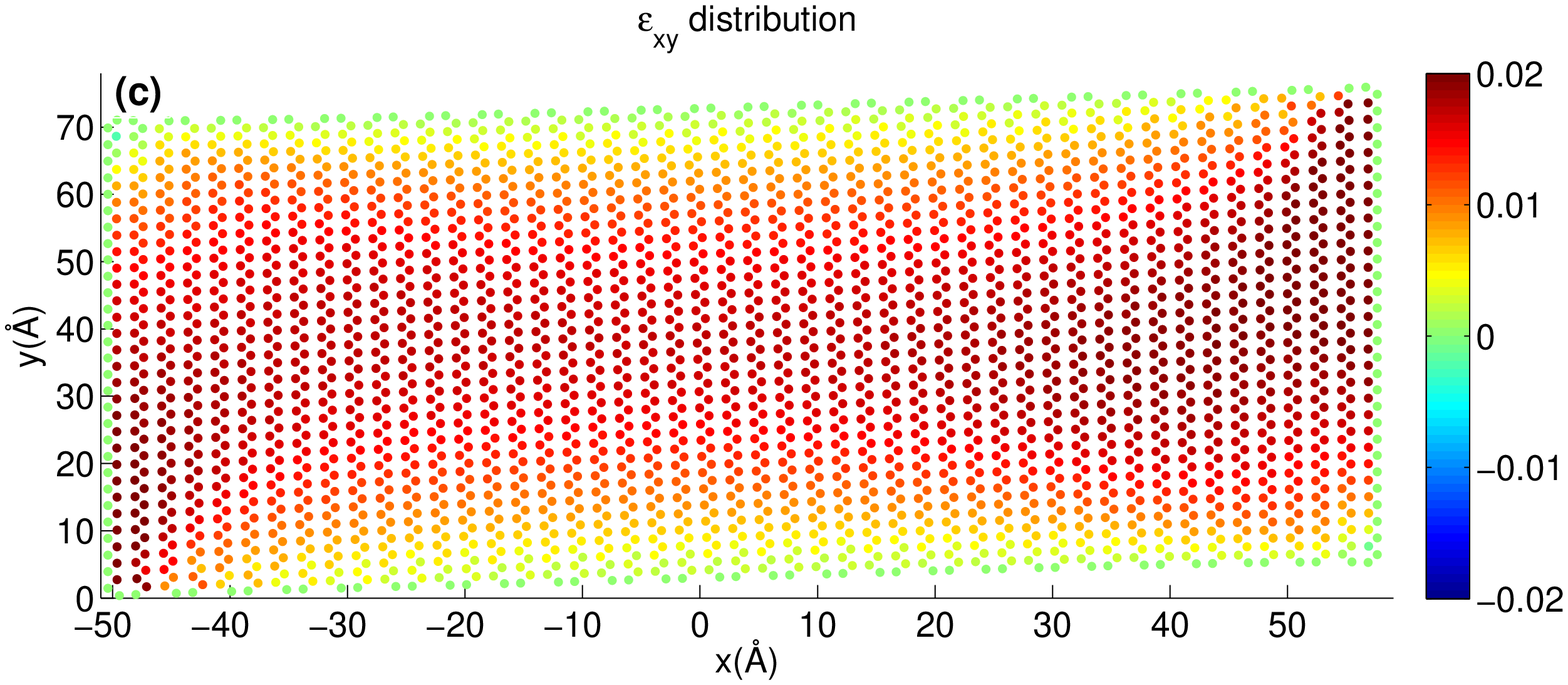}
  \caption{\label{QW_GRs}(Color online.) (a) Schematic illustration of
    AGNRs, connected to two semi-infinite quantum
    wires. There are $N$ and $M$ carbon atoms in $x$ and $y$
    directions, respectively. $\theta$ is the angle between the direction of applied strain $S$ and $x$-axis.   (b) The tensile and (c) the shear component of the strain at $30^{\circ}$ when the nominal strain is $\epsilon_{ns}=10\%$.}
    \end{figure}

Most previous theoretical studies of graphene strain sensors have adopted homogeneous junctions.  However, no lattice mismatch occurs at the interfaces for homogeneous junctions, whereas the conductance in unstrained GNRs is either zero or one at the Fermi energy $E=0$ \cite{conductance-GNRs}.  Unlike most experiments in which the gate voltage is only applied to graphene samples, the Fermi energy $E$ should vary to investigate electronic transport through GNRs. For heterogenous junctions of GNRs between quantum wire contacts, transport through GNRs is mediated by the gate voltage $V_{g}$ as in previous experimental measurement on electrical properties of graphene samples \cite{graphene1,graphene2,graphene3}. These heterogenous junctions are inspired by the fact that the contacts are metallic rather than carbon in experiments \cite{graphene1,graphene2,graphene3} and the conductance of quantum wire contacts is maximal at $E=0$ because all channels are available to electronic transport.

However, lattice mismatch may exist at the interfaces of heterogeneous junctions. Here we adopt heterojunctions of armchair-edged GNRs (AGNR) between quantum wire contacts as discussed in Ref. \cite{tm2} to minimize the effect of lattice mismatch at the interfaces and investigate the effect of uniaxial plus shear strain on electronic transport as illustrated in Fig. \ref{QW_GRs}. The strain is only applied to the AGNR and impacts the hopping integrals in the AGNR.  The hamiltonian of the AGNR and contacts is described by the tight binding approximation as
\begin{equation}
\label{eq:1}
\hat{H}=\sum_{\langle ij,i^{'}j^{'}\rangle}t_{ij,i^{'}j^{'}} \hat{c}^{\dag}_{ij}\hat{c}_{i^{'}j^{'}}
+V_{g}\sum_{ij}\hat{c}^{\dag}_{ij}\hat{c}_{ij},
\end{equation}
where a pair of integers $ij$ indicates the lattice position $\vec{R}_{ij}$, and $\hat{c}_{ij}$
($\hat{c}^{\dag}_{ij}$) is the electron annihilation (creation) operator. The summation is over the nearest neighbors indicated by $\langle \cdots\rangle$. $t_{ij,i^{'}j^{'}}$ is the hopping integral between nearest-neighboring sites indexed by $ij$ and $i^{'}j^{'}$. $V_{g}$ is the effective gate voltage applied to graphene, which is zero in contacts. When $V_{g}$ slightly varies at the interfaces, the conductance does not change much around $V_{g}=0$ \cite{note1}.

The deformed configurations of the graphene nanoribbons were obtained by molecular mechanics simulations, where the strain was obtained via applied displacement loading~\cite{strain-transport4, strain-Qi02}. The rectangular AGNR consisted of 2832 atoms with a length of $L=10.224$ nm and width $W=7.018$ nm.  Displacements were applied in increments of 0.01\AA~followed by a subsequent energy minimization and relaxation until the change in system energy was less than $10^{-7}$ compared with the previous step.  The simulations were {performed} using the open source package LAMMPS~\cite{lammps}, and the AIREBO interatomic potential~\cite{potential} with a cutoff of $0.68$nm. This potential has been shown to accurately describe carbon-carbon interactions resulting in accurate predictions of the mechanical properties of graphene~\cite{MM-simulation}. We note that because molecular mechanics simulations were performed, which are intrinsically at 0K, and because all applied displacements were in-plane, there was no out-of-plane buckling during the simulation.

As shown in Fig.~\ref{QW_GRs} (b,c), coupled tension and shear were applied onto AGNRs and the corresponding strains were calculated as discussed in previous works~\cite{strain-transport4}. We also define the `nominal strain' as the displacement applied (regardless of the direction) with reference to the original length.  Once the carbon atomic positions are obtained at each value of strain, the hopping along each bond (with the length $l$) $V_{pp\pi}=t_{0}e^{-3.37(l/a-1)}$ \cite{strain-TB} ($t_{0}=2.7$ eV and $a=0.142$ nm) is used as the basis for the electronic structure and quantum transport calculations. Due to the 2D nature of our analysis, $\sigma$ bonds were not considered in our calculation.

Strain was applied with a tilted angle $\theta$ (Fig.~\ref{QW_GRs}) from $0^{\circ}$ to $90^{\circ}$ at five different angles, i.e. $0^{\circ}$, $30^{\circ}$, $45^{\circ}$, $60^{\circ}$ and $90^{\circ}$, where $0^{\circ}$ loading represents pure tension, $90^{\circ}$ represents pure shear and the other {three} couple tension and shear. To simplify the notation, we also introduce a `nominal strain' $\epsilon_{ns}$, which is defined as the displacement applied over the original nanoribbon length regardless of the loading angle.  All cases with different loading angles are deformed at three stages, namely $\epsilon_{ns} = 2.5\%, 5\%, 10\%$, and we will refer exclusively to $\epsilon_{ns}$ in the following.
The tension ($\epsilon_{xx}$) and shear ($\epsilon_{xy}$) strain components at the three stages for different loading angles are corresponding as:  $\epsilon_{xx} = 2.5\%, 5\%, 10\%$ and $\epsilon_{xy} = 0\%, 0\%, 0\%$ for $0^{\circ}$; $\epsilon_{xx} = 2.1\%, 4.1\%, 8.7\%$ and $\epsilon_{xy} = 0.5\%, 0.9\%, 1.8\%$ for $30^{\circ}$; $\epsilon_{xx} = 1.6\%, 3.3\%, 7.8\%$ and $\epsilon_{xy} = 0.7\%, 1.4\%, 2.6\%$ for $45^{\circ}$; $\epsilon_{xx} = 1.1\%, 2.4\%, 7\%$ and $\epsilon_{xy} = 0.9\%, 1.8\%, 3.1\%$ for $60^{\circ}$; $\epsilon_{xx} = 0\%, 0\%, 0\%$ and $\epsilon_{xy} = 1.1\%, 2.2\%, 3.7\%$ for $90^{\circ}$. The deformed atomic configuration and the resulting components of the tensile and shear strain at $30^{\circ}$ when the nominal strain is $\epsilon_{ns}=10\%$ is shown in Fig.~\ref{QW_GRs}(b) and (c) \cite{graphene-MM-BU}.  The left  contact is fixed while the right one is shifted with strain similar to Ref. \cite{strain-e4} when the shear strain is present, and the hopping integral in contacts and interfaces are not affected by the shear/tension.

\begin{figure}
\includegraphics[scale=0.45]{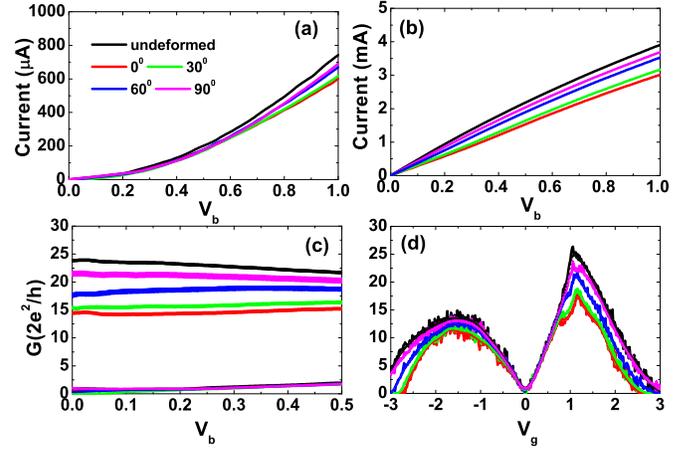}
  \caption{\label{G-Vb}(Color online.)
  The current through AGNR depends on the bias voltage $V_{b}$ at $V_{g}=0$ (a) and $V_{g}=t_{0}$ (b).
  The conductance of AGNR changes with $V_{b}$ at $V_{g}=0$ (lower lines) and $t_{0}$ (upper symbols) (c), and with the gate voltage in AGNR at $V_{b}=0$ and $V_{b}=0.2t_{0}$ (d). No strain is applied (black line) and the nominal strain $\epsilon_{ns}$ being $10\%$ is applied at $0^{\circ}$ (red line), $30^{\circ}$ (green line), $60^{\circ}$ (blue line) and $90^{\circ}$ (magenta line) respectively. The size of AGNRs is $M=29$ and $N=96$. $V_{b}$ and $V_{g}$ in figures is in the unit of $t_{0}$.}
\end{figure}

At a finite bias voltage $V_{b}$, the current transfer from the left contact to the right one is expressed as $I(S,V_{b},V_{g})=2e/h\int_{E_{f}-V_{b}/2}^{E_{f}+V_{b}/2}T(S,E,V_{g})dE$ \cite{Landauer}, where $S$ refers to the strain and $S=0$ stands for no deformation in the GNR, {e is the electron charge and $h$ is Planck's constant.} $T(S,E,V_{g})$ is the transmission at the strain $S$, the energy $E$ and the gate voltage $V_{g}$. Since the effect of the gate voltage on the strain sensor has been explored in a dual-gate setup \cite{strain-e4}, we include $V_{g}$ to estimate the stability and applicability of the graphene-based strain sensor. {Based on the tight binding hamiltonian in Eq.(1) and the transfer matrix method, the transmission $T(S,E,V_{g})$ is obtained through the scattering matrix by solving the Schr$\ddot{o}$dinger equations, which is described in detail in Refs. \cite{tm3,tm4}.}

$I(S,V_{b},V_{g})$ at $V_{g}=0$ at first increases slowly and then sharply with $V_{b}$ as shown in Fig. \ref{G-Vb} (a), since the transmission increases as the GNR deviates from being electrically neutral, which has the lowest density of states \cite{strain-transport5}. $I(S,V_{b},V_{g})$ at $V_{g}=t_0$ at first increases linearly and then sub-linearly with $V_{b}$ as shown in Fig. \ref{G-Vb} (b), since the transmission decreases as GNR deviates from the highest density of states \cite{strain-transport5}. The conductance is defined as $G(S,V_{b},V_{g})=I(S,V_{b},V_{g})/V_{b}$, and is directly related to the transmission at Fermi energy $T(S,E_{f},V_{g})$ at the limit of $V_{b}\rightarrow 0$ \cite{Landauer}. The conductance $G(S,V_{b},V_{g})$ slightly increases and decreases at $V_{g}=0$ and $V_{g}=t_0$ respectively when $V_{b}$ increases, as shown in Fig. \ref{G-Vb}(c). When $V_g$ changes, a large oscillation of the conductance at $V_{b}=0$ is induced by quantum interference when electrons are reflected at the GNR-contact interfaces \cite{strain-transport5,tm2,tm1,tm3,tm4}, and electron-hole asymmetry in conductance is originated by odd-numbered rings at the interfaces as shown in Fig. \ref{QW_GRs} \cite{tm2}. The curve of the conductance versus the gate voltage at $V_{b}=0.2t_{0}$ overlaps the one at $V_{b}=0$ and the fluctuation becomes invisible, as shown in Fig. \ref{G-Vb}(d), due to the summation of the transmission among the energy range $[E_{f}-V_{b}/2, E_{f}+V_{b}/2]$.

In strained AGNRs, the transport depends on both the direction and magnitude of the strain $S$. When the nominal strength of the strain is $10\%$, the current at both zero and finite $V_{g}$ is lower than in undeformed AGNR, and gradually increases with the change in angle of the applied strain, when the strain varies from $0^{\circ}$ to $90^{\circ}$ combined with shear, as shown in Fig. \ref{G-Vb}(a,b). Under pure tension at $0^{\circ}$, the change in the current completely reproduces the observation \cite{strain-e1}. The slope of the $I$-$V_{b}$ curve (i.e., the conductance) for nearly electrical neutral graphene samples, which corresponds to our result at $V_{g}=0$ and which is dependent on the specific experimental setup, is $1.5\times 10^{-4}\Omega^{-1}$ \cite{strain-e1} and $2\times 10^{-5}\Omega^{-1}$ \cite{strain-e2}. Our results show that the ballistic conductance of the 7.018 nm wide AGNR is around $3\times 10^{-5}\Omega^{-1}$ and $1\times 10^{-3}\Omega^{-1}$ at $V_{g}=0$ and $V_{g}=t_{0}$ respectively. The conductance dependence on the strain shown in Fig. \ref{G-Vb}(c) is the same as the current dependence on the strain.

We compare the conductance dependence on pure tension and pure shear at $90^{\circ}$ to estimate the effect of tension and shear on transport through AGNR. On one hand, in contrast to the fact that the conductance at $-t_{0}<V_{g}<0.5t_{0}$ under pure tension at $90^{\circ}$ is higher than undeformed AGNRs as a result of an increase in the hopping integrals along the horizontal direction \cite{strain-transport5}, the conductance under pure shear at $90^{\circ}$ is always lower than that in undeformed AGNRs as $|V_{g}|\ge 0.5t_{0}$ except for some fluctuations as shown in Fig. \ref{G-Vb}. Our calculation is consistent with the observation of the conductance dependence on the shear strain \cite{strain-e4}. On the other hand, the conductance around the neutral point (i.e., at small $|V_{g}|$) increases with the angle of tension/shear as shown in Fig. \ref{G-Vb}(c), compared with that under pure tension at $0^{\circ}$. The maximal conductance of undeformed AGNR being $MG_{0}$ ($G_{0}=2e^{2}/h$) occurs at $E=\pm t_{0}$ and $V_{g}=0$ between graphite contacts and at $E=0$ and $V_{g}=t_{0}$ between quantum wire contacts. The maximal conductance of strained AGNR around $V_{g}=t_0$ decreases as a result of deformation in AGNR when tension and/or shear is applied to AGNR. The data indicates that electronic transport through AGNR can be easily mediated by the strain when the system deviates from the electrically neutral region.

\begin{figure}
\includegraphics[scale=0.45]{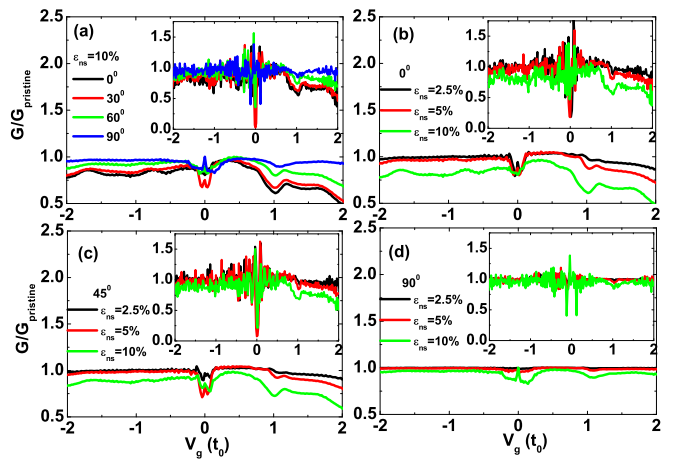}
  \caption{\label{AGNR-strain-Vb0}(Color online.)
  The ratio of the conductance in AGNRs before and after application of the strain with
  the nominal strain being 10\% (a) and the nominal strain is 2.5\%, 5\% and 10\% at an angle
  of $0^{\circ}$ (b), $45^{\circ}$ (c) and $90^{\circ}$ (d) between $x$-axis.
  The size of AGNR is $M=29$ and $N=96$. The bias voltage is $V_{b}=0.2t_{0}$ (main panels) and $V_{b}=0$ (insets).
  }
\end{figure}

In most experiments of graphene-based strain sensors \cite{strain-e1,strain-e2}, no gate voltage is applied and graphene may not be electrically neutral due to the doping from metallic contacts \cite{strain-e4,dop-contact}. Recently the effect of the gate voltage has been explored \cite{strain-e4}, and thus we set the gate voltage as a variable parameter to provide information such as the stability of the strain sensor under different gate voltages. The change of current is usually measured under strain and the percentage of the resistance is used to estimate the effect of the strain on transport through graphene samples \cite{strain-e1,strain-e2}.  Therefore, we use the ratio of the conductance compared with that in undeformed AGNR, $G(S,V_{b},V_{g})/G(0,V_{b},V_{g})$, to measure the sensitivity of the graphene-based strain sensor as shown in Fig. \ref{AGNR-strain-Vb0}.

Due to large oscillations in the conductance at $V_{b}=0$, a large oscillation is also seen in the ratio $G(S,V_{b},V_{g})/G(0,V_{b},V_{g})$ as shown in the insets of Fig. \ref{AGNR-strain-Vb0}. However, the trends of $G(S,V_{b},V_{g})/G(0,V_{b},V_{g})$ are still clear. The conductance ratio at $V_{b}=0.2t_{0}$ shown in the main panels of Fig. \ref{AGNR-strain-Vb0} is relatively smooth at negative gate voltage, shows a large dip or peak around zero gate voltage and slightly decreases as the gate voltage becomes more positive. When the nominal strain is 10\% in Fig. \ref{AGNR-strain-Vb0}(a), the conductance slightly increases but is smaller than in the undeformed case, as the angle $\theta$ varies from 0$^{\circ}$ to 90$^{\circ}$. As $\theta$ is varied from 0$^{\circ}$, 45$^{\circ}$ and 90$^{\circ}$ in Figs. \ref{AGNR-strain-Vb0}(b-d), the conductance decreases as the nominal strain increases from 2.5\% to 10\%.  It is found that the conductance shows little change under pure shear at 90$^{\circ}$ as shown in Fig. \ref{AGNR-strain-Vb0}(d). We demonstrate that this kind of strain sensor is robust since the relative proportion of the conductance is smooth within a wide gate voltage range \cite{graphene1,graphene2}.

In summary, we have studied a graphene-based strain sensor consisting of armchair graphene nanoribbon (AGNR) between metallic contacts in response to combined tension/shear. The conductance and the relative proportion of the conductance decreases as the strain increases. This kind of strain sensor has relatively higher sensitivity to the strength of the strain at finite bias voltage and a wide range of the gate voltage when the strain is parallel to the armchair edge.

Finally we comment on the performance of a strain sensor made from a zigzag graphene nanoribbon (ZGNR) between quantum wire contacts with a possible lattice mismatch at the interfaces. Compared with the case of AGNR, the fluctuation of the conductance of ZGNR is larger, when the gate voltage changes. The ratio of the conductance, $G(S,V_{b},V_{g})/G(0,V_{b},V_{g})$, ranges between 0.8 and 1.4 as $|V_{g}|\le 2t_{0}$, and the dependence of the conductance ratio on the strain are different from that seen in Fig. \ref{AGNR-strain-Vb0} when $|V_{g}|\le t_{0}$.



\emph{Acknowledgements} HSP and ZQ acknowledge support from the Mechanical Engineering
and Physics Departments at Boston University. G. P. Zhang thanks support by NSF of China (Grant
No. 11204372).

\end{document}